# USING MATHEMATICA® & MATLAB® FOR CAGD/CAD RESEARCH AND EDUCATION

§Gobithasan, R. & ◊Jamaludin M.A
§ Dept. of Mathematics, FST, KUSTEM,
Mengabang Telipot, 21030, Kuala Terengganu, M'sia.
Tel: +609-668 3132. Fax: +609-669 4660,
email: gobithasan@kustem.edu.my
§, ◊ School of Mathematical Sciences, Universiti Sains Malaysia,
11800 Minden, Penang, M'sia.
Tel: +604-653 3656. Fax: +604-657 0910,
email: jamaluma@cs.usm.my

**Abstract:** In CAGD/CAD research and education, users are involved with development of mathematical algorithms and followed by the analysis of the resultant algorithm. This process involves geometric display which can only be carried out with high end graphics display. There are many approaches practiced and one of the so-called easiest approaches is by using C/C++ programming language and OpenGL application program interface, API. There are practitioners uses C/C++ programming language to develop the algorithms and finally utilize AutoCAD for graphics display. On the other hand, high end CAD users manage to use Auto Lisp as their programming language in AutoCAD. Nevertheless, these traditional ways are definitely time consuming. This paper introduces an alternative method whereby the practitioners may maximize scientific computation programs, SCPs: Mathematica and MATLAB in the context of CAGD/CAD for research and education.

## 1. Introduction

### 1.1 CAGD: History

The foundation of designing or product development in the field of engineering and computer graphics is Computer Aided Geometric Design (CAGD). The term computer aided geometric design was formed by R.Barnhill and R.Reisenfeld in 1974 when they organized a conference on the respective topic at the University of Utah. That conference may be regarded as the founding event of the CAGD field and it resulted in widely influential proceedings. In addition to that, at the year 1984 the journal "Computer Aided Geometric Design" was founded by R.Barnhill and W.Boehm [1].

### 1.2 CAx Education

The mathematical entities of product development involve CAGD functionalities which lead to Computer Aided Design, CAD, systems development. Three general phases of product development are creative, conceptual and engineering phase. CAx (Figure 1) is an acronym which represents various IT systems support of all the phases involve in the lifecycle of a product [2].

Generally, the mathematical foundation of CAGD/CAD has been classified into the following categories [3]:

- Basic mathematics: Linear algebra, vector algebra, transformations, basic analytic geometry, equations (algebra, ordinary differential equations, partial differential equations), calculus, etc.

- Advanced mathematics: Analytic curves and surfaces, basic differential geometries, basic optimization techniques.

- Advanced CAD topics: Non-Uniform Rational B-Spline (NURBS) curves and surfaces, Boundary representations (B-reps) and Constructive Solid Geometry (CSG) techniques, Intersections (curve/curve, curve/surface, surface/surface) and Boolean Operators.

- Other advanced CAGD/CAD topics, such as: Non-linear equation solvers, Constraint solvers, Shape interrogations (e.g. curvature maps, contouring, offsets, geodesics, zebra strips, reflection lines and etc.).

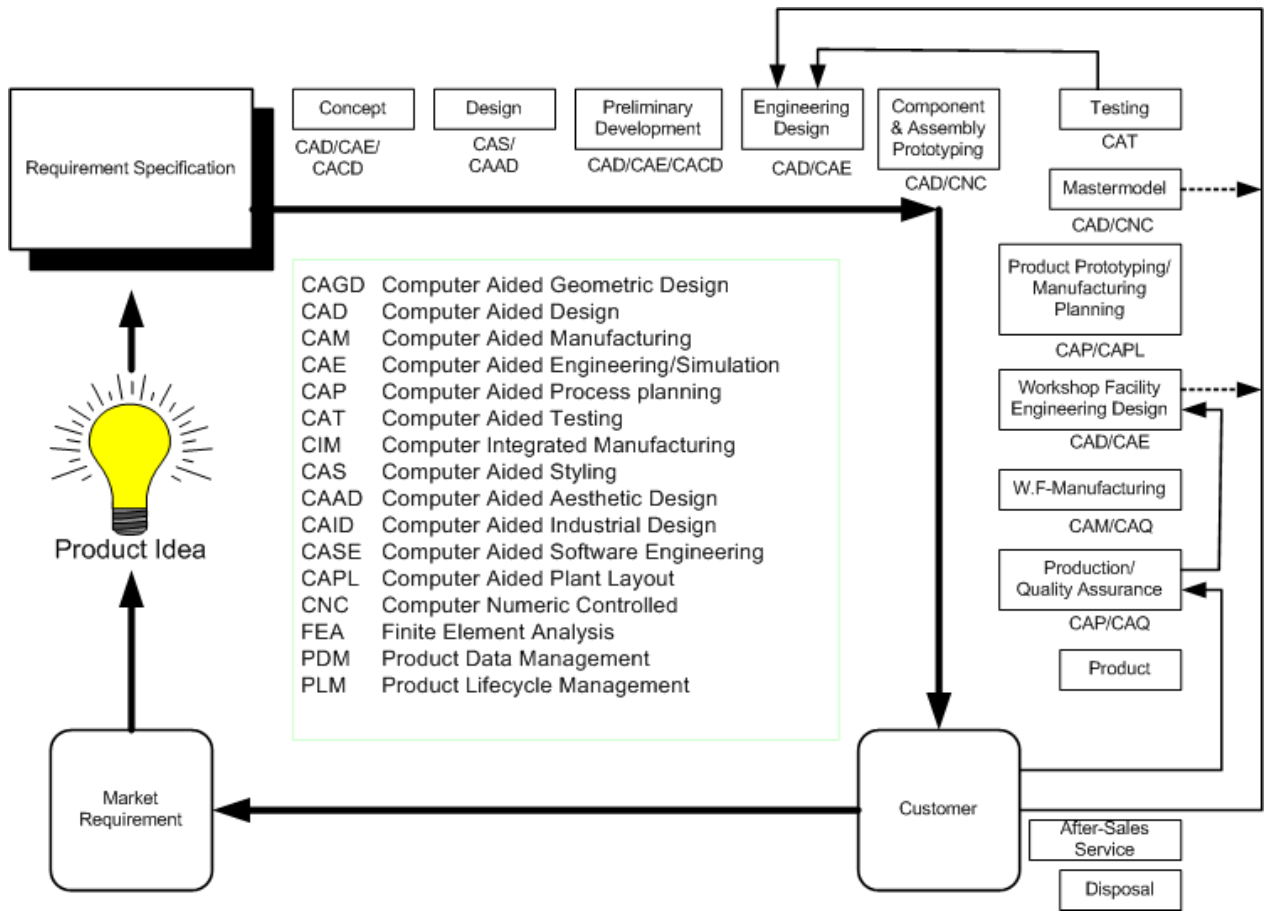

Figure 1: CAx-systems and its general process chain of product development [2].

## 2. Scientific Computation Programs

In recent years, there have been amazing advances in computer hardware and software technologies. The two categories of programming languages are high level programming and low level programming. A low level programming language is a language that provides little or no abstraction from a computer's microprocessor and it can produce efficient code. An example of low level programming language is assembly language. In high level programming, complex programming can be broken up into simpler elements and keeping programmers from having to "reinvent the wheel". C++, C, FORTRAN and Java are few examples of high level programming languages.

Scientific Computation Programs, SCPs, are new types of programs which is available in the market. These types of programs can be divided into two groups; either it functions as symbolic computation (Mathematica® and Maple®) or numerical computation (MATLAB®).

The SCPs are said more efficient because [4]:

- many mathematical and programming commands and libraries are incorporated
- the embedded algorithms are tested and optimized
- it has user friendly features and incredible graphical capabilities and
- it has procedural, functional and object-oriented programming features

With the stated advantages, the SCPs, specifically Mathematica® [5] and MATLAB® [6] are not just utilized for educational purposes, but also by researchers in the industrial sector.

## 3. Mathematica®

### 3.1 Overview

Mathematica® is a product by Wolfram Research Incorporated and it is an interactive SCP for doing mathematical computation. It can handle symbolic and numerical calculations and it is integrated with graphics system, documentation system and connectivity to other applications [6]. User may define his own procedure as it is incorporated with high level programming language which allows the user to define his own procedures.

### 3.2 Its capabilities in visualization

Mathematica® can deliver integrated and affordable high-end dynamic visualization and simulation capabilities. With nVizx™ [7], it enables users to take large volumes of natural and synthetic data, add complex calculations and generate high-resolution, dynamic images for analysis and presentation. Its capabilities in visualization are [8]: transparency, texturing, interactive 3D models, data driven real-time 3D animations, real-time control of animations, full motion camera functions, true parametric curves and surfaces plots, real-time surface approximations, fast visualization of large files, powerful animation scripting tools and etc.

### 3.3 Examples

In this section, four selected illustration are shown with regards of CAGD (Detail notes of CAGD/CAD can be found in [9,10]).

The first example (figure 2) shows an educational example of a way to represent a circle using open quadratic NURBS. To note, the outer closed curve is a B-Spline curve. The points indicate its control points. By implementing de Boor algorithm, only a page of Mathematica® programming needed to generate figure 2.

The second example (figure 3) shows a research example: a Bezier spiral which has been developed by Walton & Meek [11]. One may visually analyze the curve by few lines of programming in Mathematica®. The points indicate end points of Bezier cubic and the circle indicates the circle of end curvatures. Bezier spiral is considered high end curve where it can be used for highway and robot trajectory design.

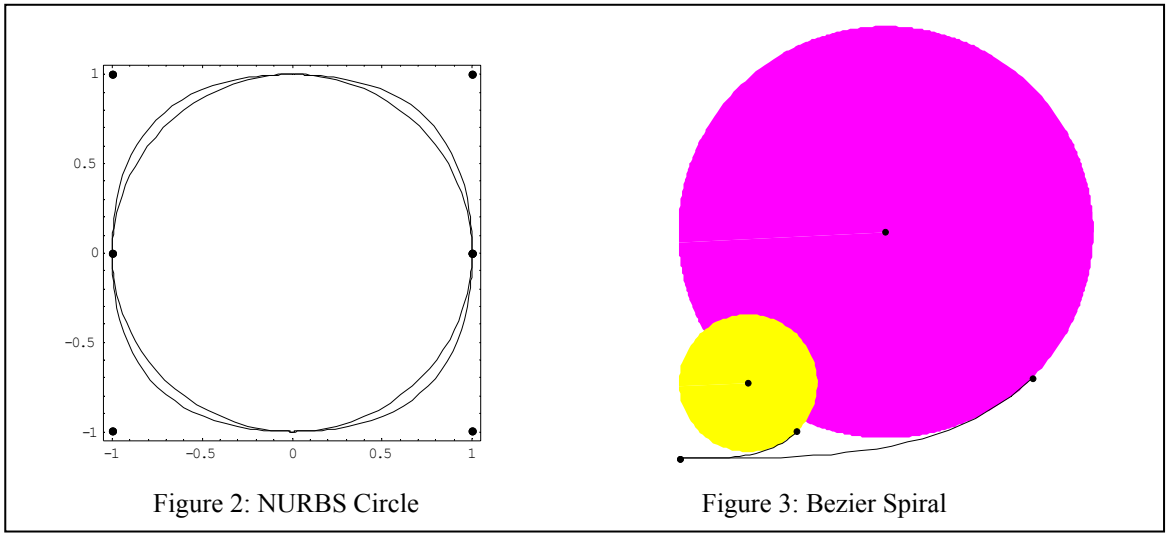

Figure 2: NURBS Circle                    Figure 3: Bezier Spiral

Figure 4 illustrates an example of a goblet generated from a profile of a $G^2$ continuous Timmer curve. Mathematica® 5.0 was utilized to generate a general algorithm to construct of $G^2$ Timmer curves [12,13,14]. To note, the symbolic computing engine in Mathematica® has been maximized during the involvement of algebraic manipulation, differentiation and etc.

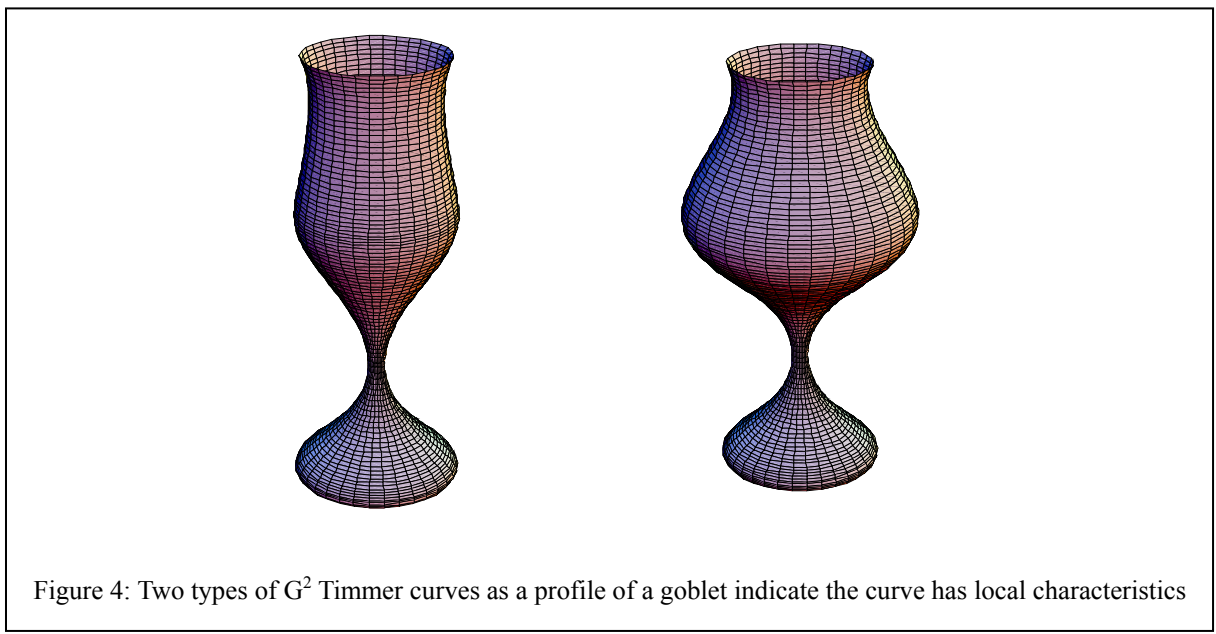

Figure 4: Two types of $G^2$ Timmer curves as a profile of a goblet indicate the curve has local characteristics

There are many ways to generate surfaces namely surface of revolution (figure 4), extrude, tensor product, ruled surface, swept surface and etc. An example of swept surface (by means of surface of revolution) using a closed profile of Bezier quartic is illustrated in figure 5 [15].

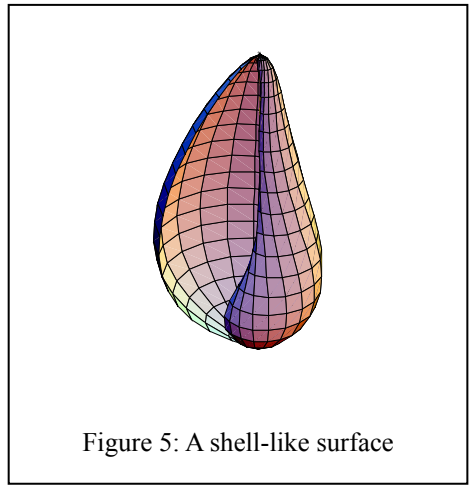

Figure 5: A shell-like surface

There is a package available in Mathematica specializing for CAGD and computer graphics developed by Andres Iglesias's group [19]. This package has been created using Mathematica programming language and it can be used along with Mathematica palettes which makes it easier for students to experiment the technicalities in the field.

## 4. MATLAB®

### 4.1 Overview

The MATLAB language was created by a specialist in numerical analysis: Prof. Cleve B. Moler, Professor of Computer Science at the University of New Mexico. It has since expanded to commercial and open-source derivatives of the original MATLAB language. Today, the premier commercial version is MATLAB® (The MathWorks is the developer and supplier of MATLAB®) and the open source version of it is Octave [16].

The name MATLAB® itself is a contraction of the words "MATrix LABoratory". It has become the de facto standard for digital signal processing and been used extensively in analysis and rapid prototyping. It is said to be more powerful and productive compared to the conventional programming languages like C and FORTRAN [6].

### 4.2 Graphics rendering in MATLAB®

MATLAB® has three types of graphics Renderers [6]: Painter, ZBuffer, and OpenGL. The original renderer in MATLAB® is Painters and it does not support lighting and transparency. ZBuffer may handle 3D graphics efficiently along with lighting capabilities. On the other hand, the OpenGL is actually application program interface (API) and is faster than ZBuffer and Painter. The OpenGL interface provides implementation either at software or hardware level and it also provides object transparency, lighting and accelerated performance. Detail information on controlling the Renderers is available from [17].

### 4.2 Examples

This section illustrates some examples of CAGD technicalities with regards of MATLAB®. The first example indicates how a designer utilizes Cell Mode (available in MATLAB® 7.0) to interactively generate a Bezier Spiral by adjusting the control points (stars). The perpendicular lines protruding from the Bezier curve indicates its porcupine plot. This is an example of an experimental work to develop a fair curve which can be utilized for product design.

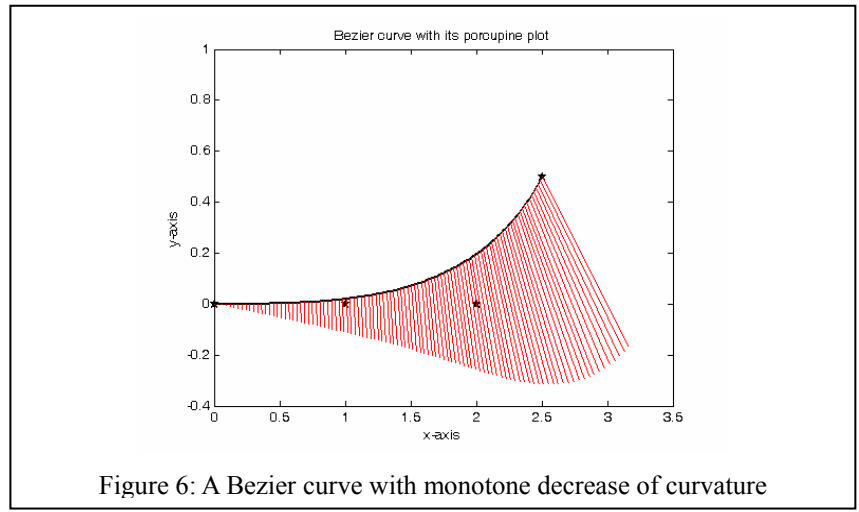

Figure 6: A Bezier curve with monotone decrease of curvature

The next example illustrates a curve called involute spline which is recently developed especially for product design [18]. This type of spline is can easily connected with $G^2$ continuity and has the characteristic of high end curve. MATLAB® has been used to simulate and analyze the curve (figure 7).

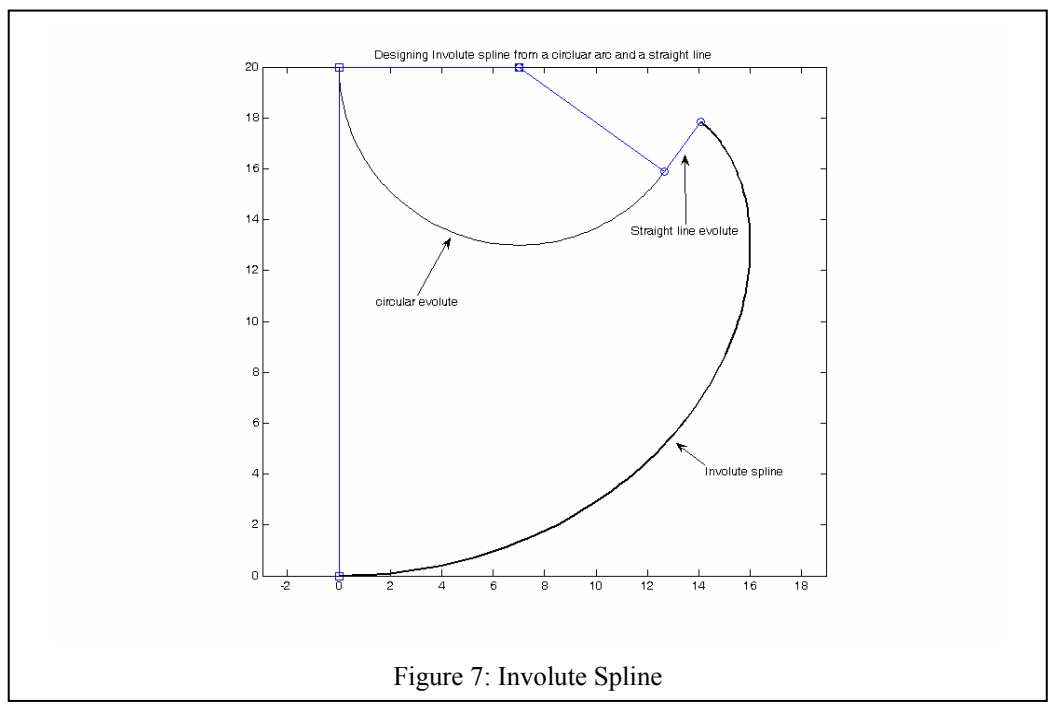

Figure 7: Involute Spline

Finally, the last example illustrates an example of designing Bezier surface (tensor product method) by using a package developed by Andres Iglesias [4]. We have modified the package in order to visualize isolines on the surface for shape interrogation purposes. Thus, the modified package can be used to design Class A surfaces. The circles indicate its control points.

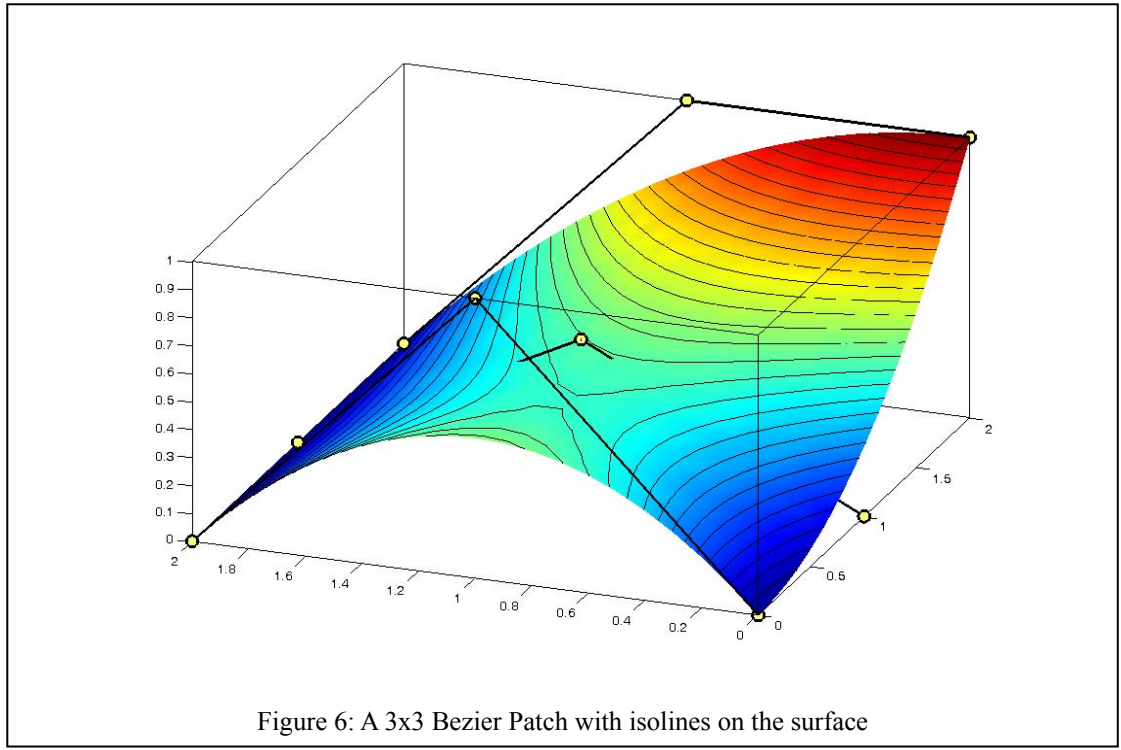

Figure 6: A 3x3 Bezier Patch with isolines on the surface

## 5. Conclusion

In this paper, we have shown how Mathematica® and MATLAB® can be utilized in the context of research and education of CAGD/CAD. Since geometric information are stored in a standardized formats, like IGES, DXF, CATIA and etc, there is a need for these packages to support/produce geometric entities in the mentioned formats. Early work can be seen in [19] where the researchers have successfully developed an IGES-Mathematica converter. Thus by producing such a converter in Malaysia, Small Medium Industries, SMIs, may use the package for industrial use.